\pdfoutput=1

\documentclass[11pt,table]{article} 

\usepackage{acl}

\usepackage{times}
\usepackage{latexsym}
\usepackage{multirow}

\usepackage[inline,shortlabels]{enumitem}
\setlist[itemize]{noitemsep}
\usepackage[T1]{fontenc}
\usepackage{soul}
\usepackage{float}

\usepackage[utf8]{inputenc}
\usepackage{amssymb}
\usepackage{amsmath}
\usepackage{booktabs}
\usepackage{enumitem}
\usepackage{graphicx} 
\usepackage{subcaption}
\usepackage{pbox}

\usepackage{microtype}

\usepackage{inconsolata}

\usepackage{graphicx}

\newcommand{\B}[1]{{\textbf{#1}}}
\newcommand{\U}[1]{{\underline{#1}}}
\usepackage{tcolorbox}
\definecolor{lightgray}{gray}{0.9} 

\NewDocumentCommand{\emark}{}{%
  \text{\normalfont\includegraphics[height=1.3\fontcharht\font`0]{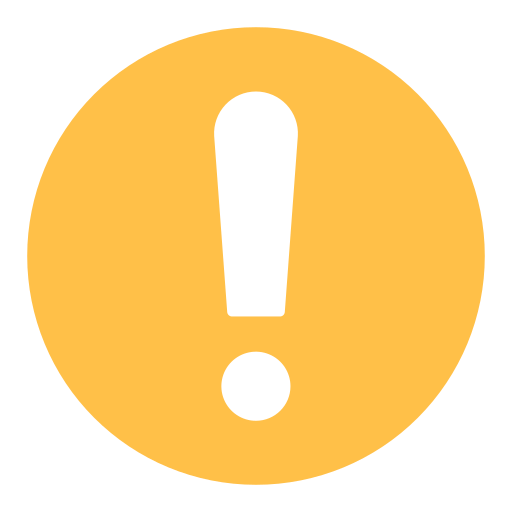}}%
}
\NewDocumentCommand{\qmark}{}{%
  \text{\normalfont\includegraphics[height=1.3\fontcharht\font`0]{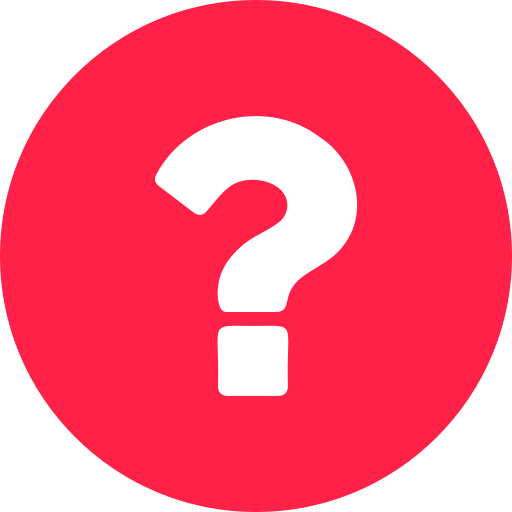}}%
}
\NewDocumentCommand{\cmark}{}{%
  \text{\normalfont\includegraphics[height=1.3\fontcharht\font`0]{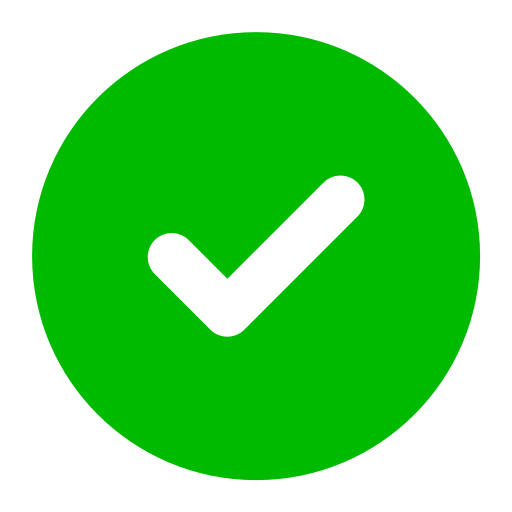}}%
}

\tcbset{
  myprompt/.style={
    colback=gray!5,       
    colframe=gray!50,      
    fonttitle=\bfseries,   
    coltitle=black,        
    boxrule=0.5mm,         
    width=\textwidth,      
    top=5pt, bottom=1pt,   
  }
}

\author{
    Jia-Huei Ju$^1$\; Suzan Verberne$^2$\; Maarten de Rijke$^1$\; Andrew Yates$^3$ \\
    $^1$University of Amsterdam\; $^2$Leiden University \\ $^3$Johns Hopkins University, HLTCOE \\
    \texttt{\{j.ju, m.derijke\}@uva.nl},\;
    \texttt{s.verberne@liacs.leidenuniv.nl},\\
    \texttt{andrew.yates@jhu.edu}
}

\title{Controlled Retrieval-augmented Context Evaluation for Long-form RAG}

\begin{document}
\maketitle
\begin{abstract}
Retrieval-augmented generation (RAG) enhances large language models by incorporating context retrieved from external knowledge sources. While the effectiveness of the retrieval module is typically evaluated with relevance-based ranking metrics, such metrics may be insufficient to reflect the retrieval's impact on the final RAG result, especially in long-form generation scenarios. We argue that providing a comprehensive retrieval-augmented context is important for long-form RAG tasks like report generation and propose metrics for assessing the context independent of generation. We introduce CRUX, a \textbf{C}ontrolled \textbf{R}etrieval-a\textbf{U}gmented conte\textbf{X}t evaluation framework designed to directly assess retrieval-augmented contexts. This framework uses human-written summaries to control the information scope of knowledge, enabling us to measure how well the context covers information essential for long-form generation. CRUX uses question-based evaluation to assess RAG's retrieval in a fine-grained manner. Empirical results show that CRUX offers more reflective and diagnostic evaluation. Our findings also reveal substantial room for improvement in current retrieval methods, pointing to promising directions for advancing RAG's retrieval. Our data and code are publicly available to support and advance future research on retrieval for RAG.\footnote{\url{https://github.com/DylanJoo/crux}}
\end{abstract}

\section{Introduction}
With their emerging instruction-following capabilities~\cite{Ouyang2022-ps, Wei2021-xk}, large language models (LLMs) have adopted retrieval-augmented generation (RAG)~\cite{Lewis2020-nf, Guu2020-tg} to tackle more challenging tasks, such as ambiguous question answering (QA)~\cite{Stelmakh2022-kt, Gao2023-dc} and long-form response generation~\cite{Shao2024-ef}.
The role of retrieval in RAG is to access information from external sources and prompt it as plug-in knowledge for LLMs. 
To achieve this, typical RAG systems retrieve the $k$ most relevant chunks as the retrieval-augmented context (abbreviated as \emph{retrieval context}, hereafter), and prompt the LLM to generate a response using this information.

\begin{figure}
    \centering
        \includegraphics[width=\linewidth]{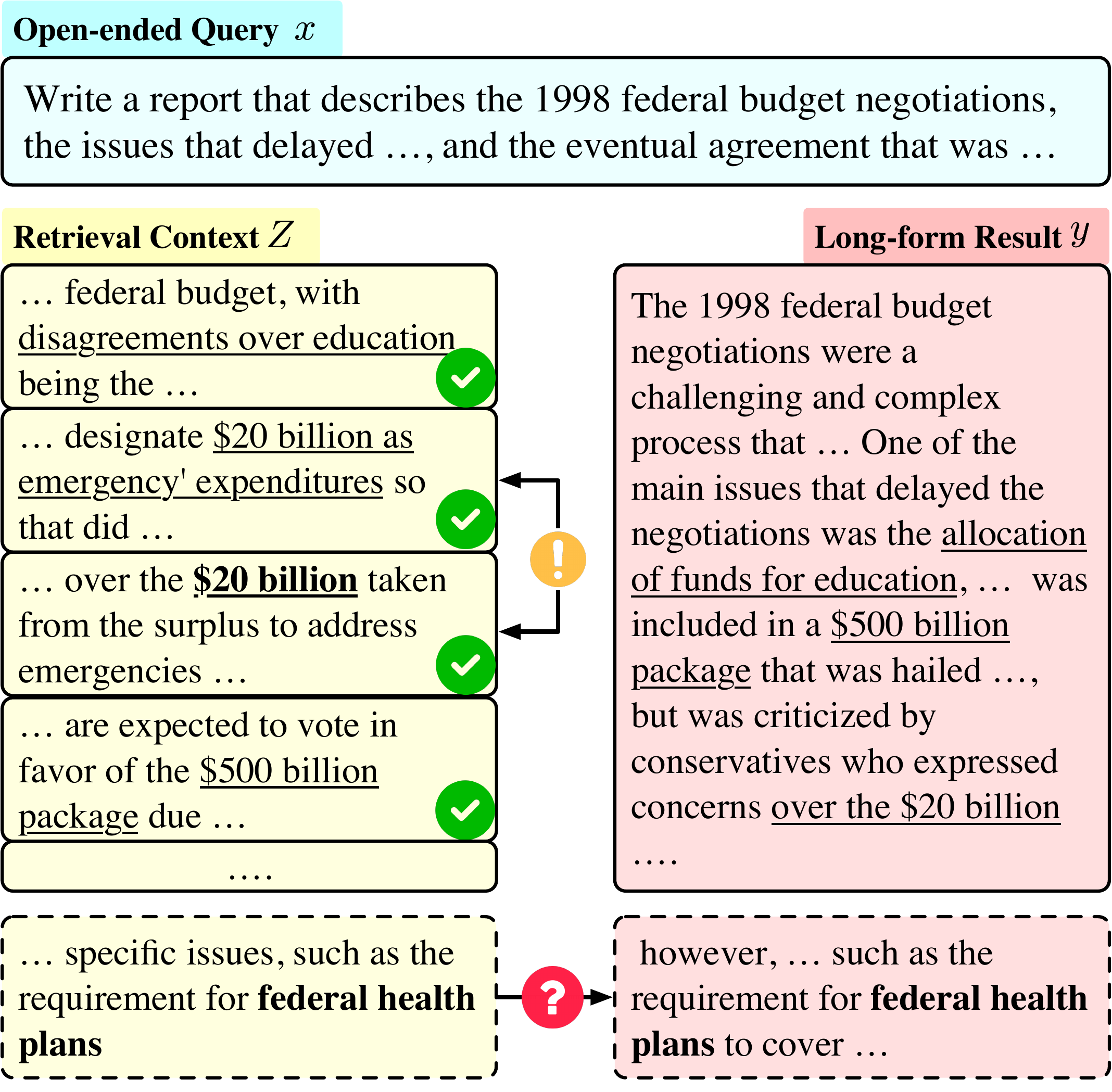}
        \caption{An example of long-form generation with an open-ended query $x$ and a desired response $y$.
        The underlined text marks relevant content in the retrieval~(\cmark) that contributes to the final result. 
        By directly assessing the retrieval context $Z$, we can further explicitly identify incomplete~(\qmark) and redundant retrieval~(\emark).}
        \label{fig:teaser}
\end{figure}

A suboptimal retrieval context hinders the generation process~\cite{Asai2024-ow, Rau2024-lk}, triggering negative impacts and resulting in unsatisfying final RAG results.
One widely-studied effect is the impact of noise from irrelevant retrieval~\cite{Yoran2023-me}, which increases the risk of hallucinations~\cite{Asai2022-tq} and distractions~\cite{Shi2023-dc}.
Such prior studies have mainly focused on short-answer tasks; however, recent RAG research has shifted towards generating comprehensive and structured reports with open-ended queries~\cite{Zhao2024-mn, Lawrie2024-lv}, as illustrated in Figure~\ref{fig:teaser}, introducing new concerns of suboptimal retrieval.

In the scenario of open-ended queries where a short answer is insufficient and a long-form result is required, incompleteness and redundancy emerge as the critical yet underexplored negative impacts from retrieval~\cite{Joren2024-ov}.
Specifically, (i) \textit{incomplete} retrieval fails to capture the full nuance of the query, leading to partial or misleading generations, and
(ii) \textit{redundant} retrieval contexts restrict the diversity of knowledge, undermining the usefulness of augmented knowledge~\cite{Yu2024-ab, Chen2024-zz}. 
Figure~\ref{fig:teaser} exemplifies such impacts of suboptimal retrieval matters on the final long-form RAG result.

To examine these effects, a suitable retrieval evaluation framework is crucial for measuring completeness and redundancy in the retrieval context.
Current retrieval evaluation practices are insufficient for measuring retrieval effectiveness in long-form RAG, as they are designed for web search~\cite{Bajaj2016-on} or short-answer QA~\cite{Kwiatkowski2019-uw}. They only require a focus on relevance-based ranking, which can be simply evaluated with retrieval metrics such as MRR and Recall@$k$.
In contrast, long-form RAG requires retrieving multiple aspects and subtopics to ensure completeness, which goes beyond surface-level relevance~\cite{Tan2024-ff, Grusky2018-ss}.

To address the gap, we propose a \textbf{C}ontrollable \textbf{R}etrieval-a\textbf{U}gmented conte\textbf{X}t evaluation framework (CRUX).
The framework includes controlled evaluation datasets and uses coverage-based metrics that directly assess the content of the retrieval context instead of relevance-based ranking.
We use human-written multi-document summaries to define the scope of the retrieval context, enabling a controlled oracle retrieval for more diagnostic evaluation results.
Finally, we assess both the (intermediate) retrieval context and (final) RAG result via question-based evaluation~\cite{Sander2021-lg,Dietz2024-iq}, supporting fine-grained evaluation between them.

To validate the usability of our evaluation framework, we conduct empirical experiments with multiple retrieval and re-ranking strategies, including relevance and diversity re-ranking.
Empirical results demonstrate the limitations of suboptimal retrieval in terms of coverage and density.
Our additional metric analysis further demonstrates that relevance ranking metrics lack coverage-awareness, highlighting CRUX’s strength in identifying retrieval impacts on long-form RAG.
Notably, our framework balances scalability and reliability by integrating LLM-based judgments with human-grounded data.
Our final human evaluation also confirms CRUX's alignment with human perception.

Overall, our controlled retrieval context evaluation aims to identify suboptimal retrieval for long-form RAG scenario. Our contributions are as follows:
\begin{itemize}[leftmargin=*,nosep]
    \item We create a controlled dataset tailored for evaluating the retrieval context for long-form RAG;
    \item We propose coverage-based metrics with upper bounds to help diagnosing the retrieval context in terms of completeness and redundancy;
    \item Our empirical results showcase the limitations of existing retrieval for long-form RAG;
    \item Our framework can serve as a reliable experimental testbed for developing more compatible retrieval for long-form RAG.
\end{itemize}

\section{Related Work}
\paragraph{The importance of retrieval in RAG.} 
LLMs are highly effective at parameterizing world knowledge as memory; however, accessing long-tail knowledge~\cite{Mallen2023-os} or verifying facts~\cite{Mishra2024-ov, Min2023-tz} often requires retrieving information from external sources.
This highlights the essential role of retrieval in augmenting reliable knowledge for downstream applications~\cite{Zhang2024-je, Zhu2024-kz, Rau2024-lk}, which is especially important in long-form generation~\cite{Gao2023-dc, Mayfield2024-vw, Tan2024-ff}.
Many studies point out that the limitations of retrieval lead to unsatisfying RAG results~\cite{BehnamGhader2023-ll, Su2024-sq, Asai2024-ow, Rau2024-lk}, raising the critical question: \emph{how effectively can retrieval augment knowledge for LLMs?}

\paragraph{Automatic evaluators for NLP tasks.}
LLMs have shown promising instruction-following capability, making them increasingly common as automatic evaluators across various NLP tasks~\cite{Thakur2025-cj, Zheng2023-xr, Chiang2023-pr}. 
Due to their cost efficiency and scalability, LLM-based evaluations have also been applied to information retrieval (IR)~\cite{Thomas2024-pj, Dietz2024-iq} and short-form generation tasks~\cite{Saad-Falcon2024-qz, Shahul2023-vp}. 
Instead of short-form RAG, we target long-form generation with open-ended query, which requires retrieval to ensure completeness in addition to surface-level relevance. 
Reference-based metrics like ROUGE used in summarization also fall short in such scenarios~\cite{Krishna2021-sf}.
Thus, a flexible framework is needed to assess information completeness and redundancy in the retrieval context.

\paragraph{Evaluating retrieval for long-form generation.}
Evaluation methodologies in IR and NLP have been standardized and developed for decades~\cite{Voorhees2002-dw, Voorhees2004-bg}. In recent years, nugget-based (sub-topics or sub-questions) evaluation~\cite{Pavlu2012-ch, Clarke2008-cn, Dang2008-wa} has resurfaced as an important focus due to the feasibility of automatic judgments.
Similarly, question-based evaluation that estimates the answerability~\cite{Eyal2019-rj, Sander2021-lg} of a given text is well-aligned with LLMs while preserving aspect-level granularity, making it particularly good for evaluating long-form generation.
This helps inform the development of a unified evaluation setup for both the intermediate retrieval context and final long-form results, thereby facilitating more informative evaluation for RAG's retrieval methods.

\section{Controlled Retrieval-augmented Context Evaluation (CRUX)}
This section introduces CRUX, a controlled evaluation framework for assessing the retrieval context in long-form RAG. It comprises:
(1) definitions of \emph{retrieval context} and its sub-question \emph{answerability} (\S~\ref{sec:3-1});
(2) curated evaluation datasets (\S~\ref{sec:3-2}) and 
(3) \emph{answerability}-driven performance metrics: coverage and density (\S~\ref{sec:3-3}).

\subsection{Retrieval-augmented Context}\label{sec:3-1}
Here we focus on the retrieval context as the important bottleneck in the long-form RAG pipeline. 
Formally, given an open-ended query $x$, a typical RAG pipeline is defined as:
\begin{equation}
    y \leftarrow G(x, Z, I), \quad Z \leftarrow RA_\theta(x, \mathcal{K}). \label{eq:rag}\\ 
\end{equation}
$RA_\theta$ denotes the retrieval modules that augment the retrieval context $Z$ from an external knowledge source $\mathcal{K}$ (i.e., a corpus), and $G$ is a LLM generator that takes as input a query $x$, retrieval context $Z$, and task-specific instruction prompt $I$ to generate the final long-form RAG result $y$.
Particularly, we argue that the quality of the retrieval context is a key limitation for achieving optimal RAG results and propose an evaluation framework for it.

\paragraph{Answerability measured by sub-questions.}
To assess the retrieval context quality beyond relevance-based ranking, we adopt question-based evaluation~\cite{Eyal2019-rj, Sander2021-lg}. 
We assess the content of an arbitrary text $z$ with a diverse set of knowledge-intensive sub-questions $Q=\{q_1, q_2, \ldots, q_n\}$. 
Such diversity enables these questions to serve as a surrogate for evaluating multiple aspects of a query, thereby facilitating explicit diagnosis of underlying concerns such as completeness and redundancy.
Specifically, we use an LLM to judge how well the text $z$ answers each sub-question and estimate a binary sub-question \emph{answerability} value (answerability, hereafter):
\begin{equation}
    G(z, q_i, I_{g}) \geq \eta \quad \forall q_i\in Q, \label{eq:answerability}
\end{equation}
where $I_{g}$ is a grading instruction prompt similar to the rubrics proposed by~\citet{Dietz2024-iq}. The output graded rating is on a scale of 0 to 5 (the prompt is included in Figure~\ref{fig:prompt-judge} in the Appendix~\ref{sec:appendix-1}).
$\eta$ is a predefined threshold determining whether the given text-question pair is answerable. The threshold analysis is reported in Section~\ref{sec:threshold}.

\begin{figure}
    \centering
    \includegraphics[width=\linewidth]{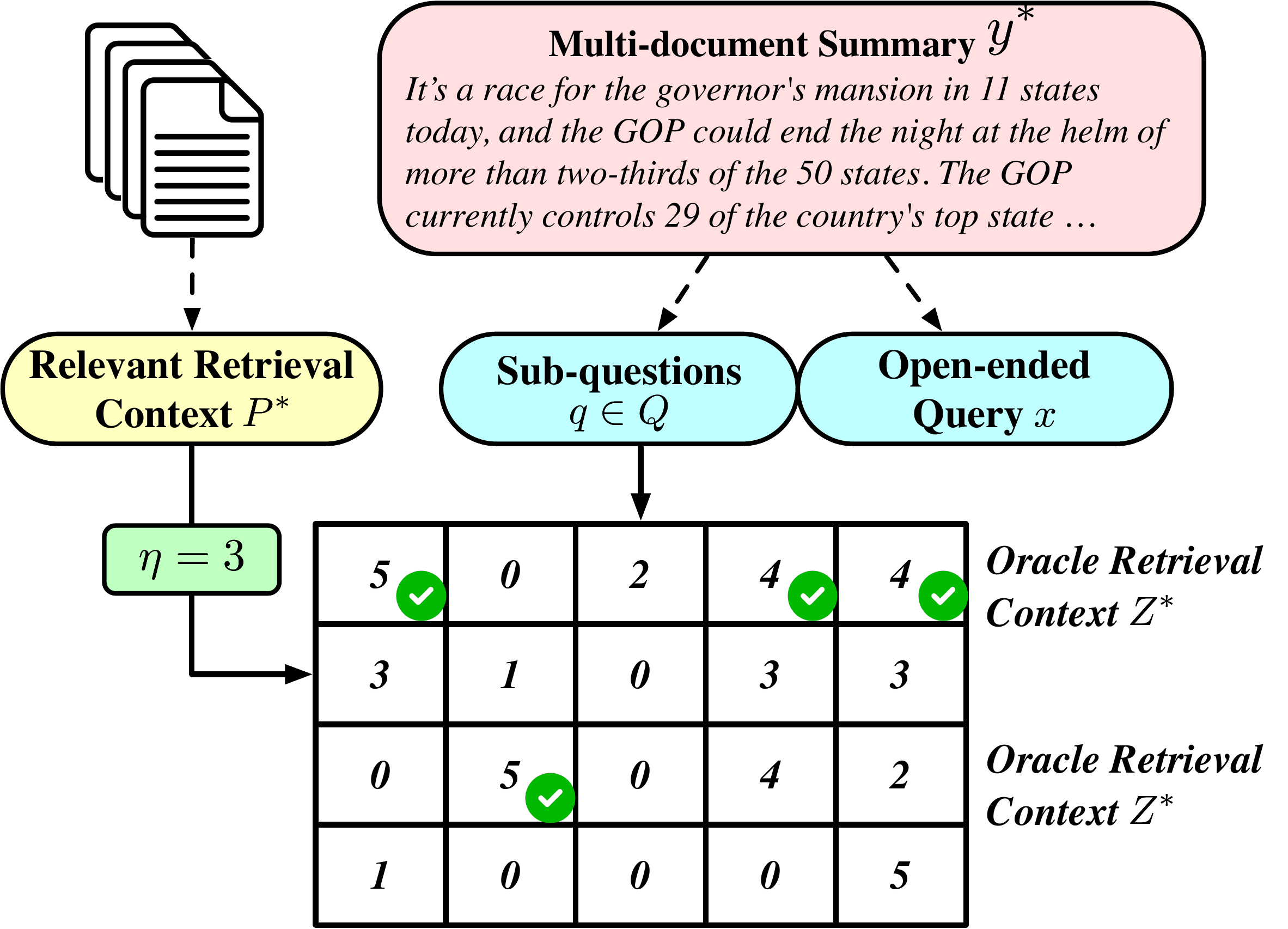}
    \caption{The controlled data generation derived from multi-document summarization datasets.} \label{fig:dgp}
\end{figure}

\subsection{Data Creation for Controlled Evaluation} \label{sec:3-2}
We further construct datasets tailored for our evaluation framework to support controlled analysis. 
As illustrated in Figure~\ref{fig:dgp}, we treat human-written multi-document summaries as the central anchor for defining:
(1) the explicit scope of the relevant retrieval context $Z^*$;
(2) an open-ended query $x$;
(3) a diverse set of sub-questions $Q$.
Together, these components support our assessment of completeness and redundancy.

\paragraph{Explicit scope of the retrieval context.}
The controllability comes from the intrinsic relationships within the multi-document summarization datasets: Multi-News~\cite{Fabbri2019-hc} and DUC~\cite{Over2004-yb}, where each example consists of a human-written summary and the corresponding multiple documents.
As illustrated in Figure~\ref{fig:dgp}, we consider the human-written summary as the proxy of an oracle long-form RAG result;\footnote{We assume the human-written summary satisfies complex information needs in the most precise and concise manner.} it is denoted as $y^*$. 
The corresponding documents $D^*$ are naturally regarded as relevant, while the other documents can be safely considered as irrelevant, forming an explicit scope for each example.
In addition, we decontextualize a document into passage-level chunks with an LLM, obtaining the set of relevant passages $p \in P^* \subseteq D^*$.
Decontextualization provides several advantages~\cite{Choi2021-yo}, ensuring the passages fit the token length limitation of all retrievers and are standalone while preserving main topics. 
Such units also help us identifying redundancy and incompleteness; see Table~\ref{tab:example} for an example.

\paragraph{Open-ended queries.}
We use an LLM to synthesize a query with open-ended information needs from the human-written summary $y^*$ via in-context prompting~\cite{Brown2020-at}. Examples are shown in Figures~\ref{fig:teaser} and~\ref{fig:prompt-tg}.
We denote these queries as $x$ in Eq.~\eqref{eq:rag}, which is the initial input for both retrieval and generation.
Such queries help expose limitations in existing retrieval systems, which often return either irrelevant or redundant passages, resulting in incomplete retrieval contexts.  
Notably, the query generation process is adaptable and can be tailored to various kinds of queries~\cite{Yang2024-xt} via similar in-context prompting. 

\paragraph{Diverse sub-questions and filtering.}
Similarly, we synthesize a diverse set of knowledge-intensive sub-questions $Q$ from the human-written summary which cover the highlights in the oracle RAG results (i.e., $y^*$). 
Thanks to the controlled settings, for each query $x$, we enumerate all possible pairs of sub-questions $q \in Q$ and relevant passages $p\in P^*$, then judge them with an LLM.
Hence, for each relevant passage, we obtain a list of graded ratings for all the sub-question as mentioned in Eq.~\eqref{eq:answerability}.
Finally, we can obtain the matrix of graded ratings as shown in Figure~\ref{fig:dgp}.
In addition, the judged ratings can serve as consistency filtering to identify unanswerable sub-questions for mitigating out-of-scope and hallucinated questions.
These pre-judged ratings can be further reused for evaluating the retrieval context, which is also released with the data.

\paragraph{Required subset of relevant passages.}
Once we have pre-judgments of all relevant passages $p\in P^*$, 
we further identify which passages are necessary and construct a smaller subset of relevant passages, denoted as $P^{**}$.
Specifically, we define this required subset as those passages that can collectively answer all sub-questions $q\in Q$.
To do so, we first rank each relevant passage according to how many questions it can answer and greedily assign each to the subset until no additional sub-questions can be answered.\footnote{The default answerability threshold $\eta$ is set to 3.} 
Those remaining are categorized as either partially or fully redundant.

\paragraph{Data statistics.}
We collected 100 open-ended queries from Multi-News~\cite{Fabbri2019-hc} and 50 queries from DUC~\cite{Over2004-yb}.
The knowledge source $\mathcal{K}$ has around 500K passages, collected from training and test splits of Multi-News and the DUC.
We generate all data using open-source~\texttt{Llama-3.1-70B-Instruct}.\footnote{\url{https://huggingface.co/meta-llama/Llama-3.1-70B-Instruct}}~\cite{Llama-Team-AI-Meta2024-cb}. 
Detailed data statistics and generation settings are reported in Appendix~\ref{sec:appendix-1}.

\subsection{Evaluation Metrics} \label{sec:3-3}
We use three metrics to assess the retrieval context for long-form RAG.
We begin by measuring a retrieval context's completeness using \emph{coverage}, then introduce derived metrics: \emph{ranked coverage} and \emph{density} to further take redundancy into account.

\begin{figure}
    \centering
    \includegraphics[width=\linewidth]{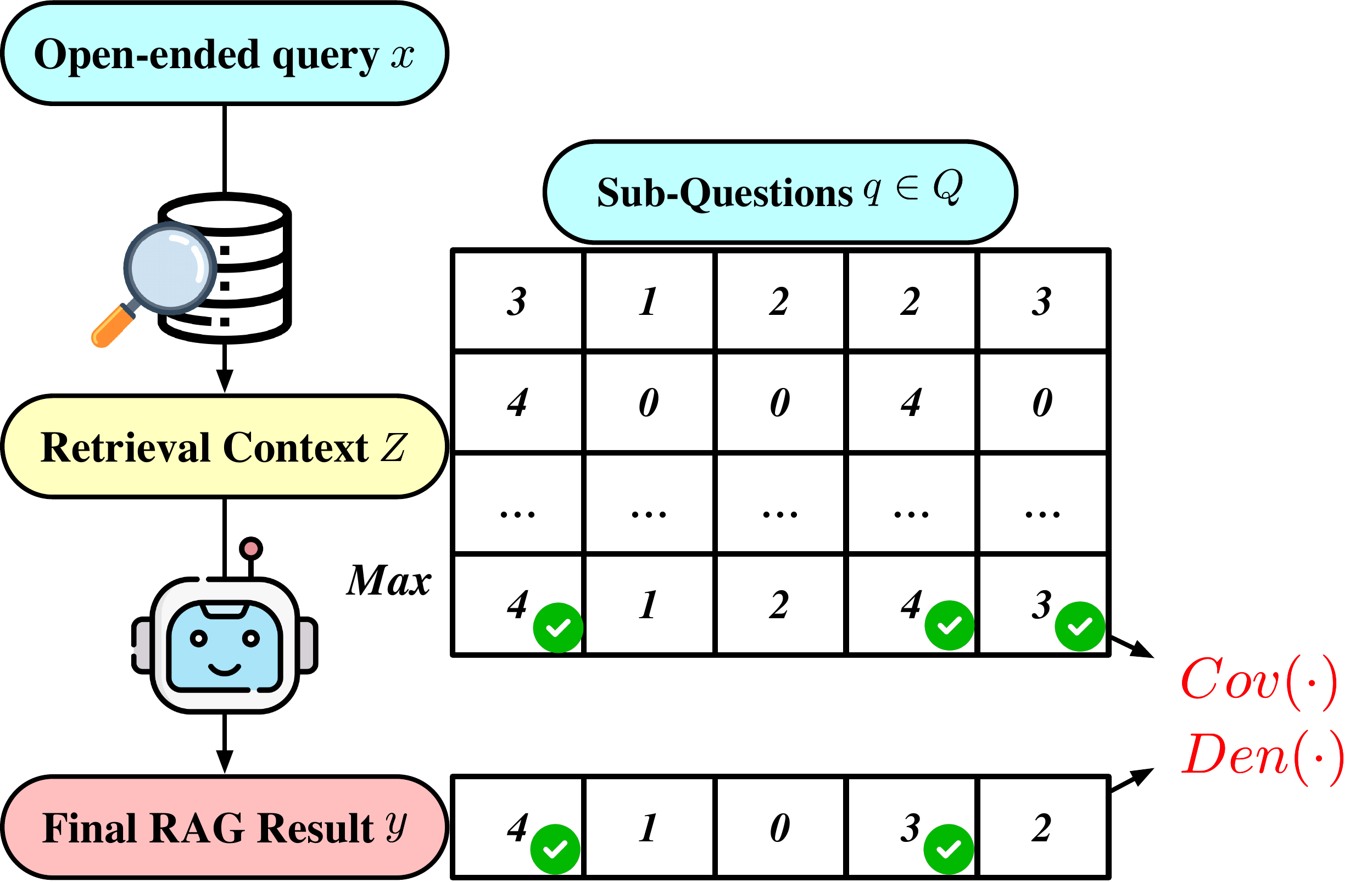}
    \caption{CRUX employs sub-question answerability to directly assess the textual content of both the retrieval context $Z$ and its corresponding RAG result $y$. The metrics include \emph{coverage} and \emph{density}.}
    \label{fig:ep}
\end{figure}

\paragraph{Coverage~$(Cov)$.}
Rather than evaluating the retrieval results based on only their relevance (e.g., nDCG and MAP), we assess the content of the retrieval contexts based on \emph{answerability}.
Given a retrieval context $Z$, we explicitly quantify the context's coverage with how many questions it can answer over the answerable sub-questions.
To compute this, we aggregate graded ratings by taking the maximum across passages in the retrieval context $Z$ and obtain binary answerability as depicted in Figure~\ref{fig:ep}. We finally normalize it by the total number of answerable sub-questions.
Formally, the coverage of the retrieval context is defined as:
\begin{align}
\mbox{}\hspace*{-2mm}
    \resizebox{0.9\hsize}{!}{$
        Cov (Z) = \dfrac{\# \{ 
            q \in Q | \max \Big( G(p\in Z, q, I_{g}) \Big) 
        \geq \eta \} }{|Q|}.
    $}
    \label{eq:coverage}
\end{align}
We can also apply this formula to evaluate the coverage of the final RAG result $y$, allowing us to compare the coverage of the retrieved passages to the coverage of the generation.

\paragraph{Ranked coverage.}
We bring coverage-awareness to the novelty ranking metric, $\alpha$-nDCG~\cite{Clarke2008-cn}. $\alpha$-nDCG evaluates novelty based on subtopics, which is naturally compatible with our framework using sub-question answerability. 
Specifically, we define the \emph{ranked coverage} by treating the answerability of sub-questions as subtopics, as follows:
\begin{align}
\mbox{}\hspace*{-2mm}
    \alpha\text{-nDCG} =&
    \sum_{r=1}^{|Z|} \frac{ng(r)}{\log(r+1)} / {\sum_{r=1}^{|Z^*|} \frac{ng^*(r)}{\log(r+1)}} \\
    ng(r) =& \sum^{|Q|}_{i=1} \mathcal{I}_{i, r} (1-\alpha)^{c_{i, r-1}}
    \label{eq:rank}
\end{align}
where $r$ is the passage rank position in the retrieval context. The function $ng$ is novelty gain, representing how much new information is covered with respect to the position $r$ and sub-questions $q_i$. 
Discount factor $\alpha$ is used for penalizing redundant sub-questions when accumulating gains.

\paragraph{Density~$(Den)$.}
We evaluate the retrieval context's density from a coverage perspective. The oracle retrieval context $Z^*$ is considered as the reference, enabling us to compute relative density based on the total number of tokens. 
The density of the retrieval context $Z$ is measured by:
\begin{equation} 
\mbox{}\hspace*{-2mm}
    \resizebox{0.9\hsize}{!}{$Den (Z) = 
    \Big(
        \dfrac{ Cov (Z) / {\rm token}(p \in Z) }{ Cov (Z^*) / {\rm token}(p \in Z^*) }
    \Big)^{w},$}
\end{equation}
where ${\rm token}(\cdot)$ means the total number of tokens, and $w$ is a weighting factor. 
We set $w$ as 0.5, assuming that the information density grows monotonically but has diminishing marginal returns when reaching the optimum.

\begin{table*}[ht]
\centering
\resizebox{.97\textwidth}{!}{%
\begin{tabular}{l 
cc >{\columncolor{lightgray}}c c >{\columncolor{lightgray}}c 
cc >{\columncolor{lightgray}}c c >{\columncolor{lightgray}}c
}
\toprule
     &
    \multicolumn{5}{c}{DUC} &  \multicolumn{5}{c}{Multi-News}  \\
    \cmidrule(lr){2-6} \cmidrule(lr){7-11}
    Retrieval Context & 
    \multicolumn{1}{c}{$Cov(Z)$} & 
    \multicolumn{1}{c}{$\alpha$-nDCG} & 
    \multicolumn{1}{c}{$Cov(y)$} & 
    \multicolumn{1}{c}{$Den(Z)$} & 
    \multicolumn{1}{c}{$Den(y)$} &
    \multicolumn{1}{c}{$Cov(Z)$} & 
    \multicolumn{1}{c}{$\alpha$-nDCG} & 
    \multicolumn{1}{c}{$Cov(y)$} & 
    \multicolumn{1}{c}{$Den(Z)$} & 
    \multicolumn{1}{c}{$Den(y)$} \\
\midrule
    (\#1) Direct prompting         & - & - & 26.7 & - & - & - & - & 21.4 & - & - \\
    (\#2) Oracle result $y^*$  & - & - & 95.3 & - & 108\phantom{.01} & - & - & 94.1 & - & 111\phantom{.01} \\
    (\#3) Oracle retrieval  $Z^*$  & 100\phantom{.01} & 80.6 & 64.6 & 100\phantom{.01} & 93.8
                                   & 100\phantom{.01} & 80.6 & 61.8 & 100\phantom{.01} & 84.7 \\
    \midrule
    BM25 (LR)           & 44.4     & 35.7     & 35.8     & \U{61.2} & \U{53.4} & 39.3     & 35.4     & 38.2     & \U{50.6} & 60.0 \\
    Contriever (DR)     & 52.1     & 45.2     & 41.7     & 70.3     & 60.5     & 43.1     & 36.6     & 36.6     & 55.4     & 58.3 \\
    SPLADE-v3 (LSR)     & 49.0     & 45.0     & 41.0     & 67.7     & 59.4     & 45.4     & 40.4     & 41.3     & 60.6     & 64.3 \\
    \midrule
    LR  + MMR           & 45.6     & 36.7     & 36.4     & 65.8     & 57.2     & 41.4     & 35.2     & 37.9     & 52.9     & 58.9 \\
    DR  + MMR           & \U{42.7} & \U{35.1} & \U{33.8} & 62.6     & 53.5     & 39.0     & \U{33.5} & \U{36.1} & 51.3     & \U{57.6} \\
    LSR + MMR           & 44.2     & 35.6     & 36.5     & 64.4     & 56.5     & \U{39.2} & 33.8     & 37.3     & 51.6     & 59.2 \\
    \midrule
    LR  + miniLM        & 49.0     & 42.5     & 38.4     & 67.9     & 57.9     & 45.3     & 39.8     & 41.2     & 58.2     & 63.0 \\
    DR  + miniLM        & 49.3     & 42.9     & 39.9     & 69.3     & 59.7     & 45.1     & 40.3     & 40.4     & 57.8     & 62.4 \\
    LSR + miniLM        & 49.4     & 42.6     & 39.2     & 69.3     & 59.2     & 45.4     & 40.3     & 40.6     & 58.0     & 62.6 \\
    \midrule
    LR  + monoT5        & 50.7     & 42.4     & 37.9     & 66.5     & 56.7     & 47.9     & 40.2     & 41.6     & 58.3     & 64.0 \\
    DR  + monoT5        & 53.2     & 44.7     & 40.7     & 70.8     & 60.0     & 45.4     & 40.0     & 40.9     & 56.6     & 62.6 \\
    LSR + monoT5        & 52.8     & 43.0     & 41.1     & 68.9     & 59.2     & 44.3     & 37.7     & 38.9     & 55.4     & 61.5 \\
    \midrule
    LR  + RankZephyr    & 51.5     & 45.9     & 40.6     & 69.9     & 59.5     & 52.9     & 47.6     & 43.9     & 65.1     & 67.7 \\
    DR  + RankZephyr    & 51.1     & 48.8     & 40.6     & 67.8     & 59.2     & 53.6     & 47.2     & 44.1     & 66.0     & 66.8 \\
    LSR + RankZephyr    & 50.4     & 45.9     & 41.2     & 67.3     & 60.0     & 54.4     & \B{49.1} & 45.8     & 67.0     & \B{69.8} \\
    \midrule
    LR  + RankFirst     & 52.0     & 46.2     & 43.9     & 70.1     & 63.4     & \B{56.0} & \B{49.1} & \B{46.4} & \B{68.0} & 69.4 \\
    DR  + RankFirst     & 53.8     & \B{49.1}     & \B{44.6} & 70.9     & 64.0     & 54.5     & 47.6     & 44.4     & 66.2     & 67.4 \\
    LSR + RankFirst     & 53.6     & 48.2     & 44.3     & 70.9     & 64.0     & 54.5     & 48.2     & 46.0     & 66.5     & 69.2 \\
    \midrule
    LR  + SetwiseFlanT5 & 49.6     & 44.2     & 42.5     & 67.8     & 61.9     & 52.1     & 44.9     & 43.2     & 63.9     & 65.5 \\
    DR  + SetwiseFlanT5 & \B{56.6} & 48.4     & 44.4     & \B{74.9} & \B{64.4} & 49.9     & 43.8     & 41.0     & 61.0     & 62.5 \\
    LSR + SetwiseFlanT5 & 51.9     & 46.0     & 43.3     & 70.1     & 62.6     & 52.0     & 47.0     & 45.1     & 65.4     & 67.8 \\
\midrule
    Rank Corr. (Kendall $\tau$) & 
    \multicolumn{1}{c}{0.676} & \multicolumn{1}{c}{0.724} & - & 
    \multicolumn{1}{c}{0.733} & - & 
    \multicolumn{1}{c}{0.838} & \multicolumn{1}{c}{0.800} & - &
    \multicolumn{1}{c}{0.810} & - \\
\bottomrule
\end{tabular}%
}
    \caption{
    Evaluation results of empirical retrieval contexts $Z$ and corresponding final results $y$ (the columns in gray) on CRUX-DUC and Multi-News. 
    Scores with bold font and underlined are the highest and lowest.
    For each dataset, columns 1 and 2 show retrieval coverage and ranked coverage; column 3 shows the final result coverage.
    The last two columns are the density of the retrieval context and final result.
    The bottom row reports the ranking correlation between the retrieval context and final results.
    }
    \label{tab:main}
\end{table*}

\section{Experiments}
To validate CRUX's evaluation capability and usability, we begin with controlled experiments with empirical retrieval contexts to enable more diagnostic retrieval evaluation.
Next, we analyze metric correlations between the retrieval contexts $Z$ and the corresponding final results $y$. 
Finally, we assess CRUX's usability through human annotations and examine other configuration impacts.

\subsection{Experimental Setups}
\paragraph{Initial retrieval.}
Our experiments employ varying cascaded retrieval pipelines to augment context from the knowledge corpus.
Given an open-ended query $x$, we first retrieve the top-100 relevant candidate passages.
Three initial retrieval approaches are considered: lexical retrieval (\textbf{LR}) with BM25,\footnote{\url{https://github.com/castorini/pyserini/}} dense retrieval (\textbf{DR}) using Contriever$^{\rm FT}$~\cite{Izacard2021-zy} and learned sparse retrieval (\textbf{LSR}) using SPLADE-v3~\cite{Lassance2024-ly}. 

\paragraph{Candidate re-ranking.}
We further re-rank the 100 candidate passages with more effective models, constructing the final retrieval context $Z$.
We experiment with varying re-ranking strategies, including pointwise re-ranking models: \textbf{miniLM}~(220M) and \textbf{monoT5} (3B).
In addition, we include state-of-the-art LLM-based listwise re-ranking models: \textbf{RankZephyr}~(7B)~\cite{Pradeep2023-ku} and \textbf{RankFirst}~(7B)~\cite{Reddy2024-lr}, as well as \textbf{Setwise} re-ranking (3B)~\cite{Zhuang2023-ii}.
Lastly, we evaluate the maximal marginal relevance (\textbf{MMR}) algorithm for diversity re-ranking to consider both relevance and diversity.\footnote{We follow~\citet{Gao2024-ax} and adopt the same pre-trained encoder for MMR:~\url{https://huggingface.co/sentence-transformers/all-mpnet-base-v2}}

\paragraph{Generation.}
Llama models~\cite{Llama-Team-AI-Meta2024-cb} with 8B parameters are used for generation. We use \texttt{vLLM}~\cite{Kwon2023-we} to accelerate the inference speed and perform batch inference. 
For fair comparisons, we adopt the same configurations for all generations. Details are provided in Appendix~\ref{sec:appendix-1}.

\paragraph{Evaluation protocol.}
As our goal is to analyze how incomplete and redundant retrieval context affects the final RAG result, we assess both the quality of retrieval context $Z$ and further investigate the relationships between them and final coverage and density: $Cov(y)$ and $Den(y)$.
Notably, the explicit scope of relevant passages allows us to reuse the pre-judgments for relevant passages as shown in Figure~\ref{fig:ep}.
Unless otherwise specified, we set the default \emph{answerability} threshold $\eta$ to 3.

\subsection{Controlled Empirical Experiments}\label{sec:4-2}
CRUX suggests explicit oracle RAG settings of retrieval context $Z^*$, thereby facilitating more indicative evaluations by controlling:
(i) the number of passages in the retrieval context (i.e., top-$k$), which is set to match the size of the oracle retrieval context, $|Z^*|$; 
(ii) the maximum generation token length, which is constrained by the match token length of the oracle retrieval, ${\rm token}(Z^*)$.\footnote{We change the prompt accordingly and truncate the maximum token length if the result exceeds.}
The following research questions guide our findings.

\paragraph{What are the reference performance bounds of the retrieval context and final RAG result?}
In the first block of Table~\ref{tab:main}, we report the performance of three reference retrieval contexts and their final RAG results:
(\#1) zero-shot \B{direct prompting};
(\#2) \B{oracle results} $y^*$ (the human-written summary);
(\#3) \B{oracle retrieval context} $Z^* \triangleq P^{**}$, which is the required subset of relevant passages given in the test collection (See Section~\ref{sec:3-2}).

Unsurprisingly, we observe the lowest coverage for RAG result without retrieval (\#1), confirming that parametric knowledge in the LLM alone is insufficient to achieve high performance.
This condition serves as the empirical lower bound of RAG.
In contrast, the oracle result using the human-written summary (\#2) achieves the highest coverage by answering over 90\% of the sub-questions. 
This implies that the generated sub-questions are answerable and validate the framework's ability to capture completeness.
The RAG result with the oracle retrieval context (\#3) yields decent coverage of 64.6 and 61.8, outperforming other empirical methods in subsequent blocks in the table. 
This demonstrates an empirical upper bound for RAG's retrieval, grounded in an oracle retrieval context $Z^*$.
Overall, CRUX provides robust bounds for reference, enabling more diagnostic evaluation of RAG's retrieval regardless of the generator.

\paragraph{How effective are empirical retrieval contexts regarding the performance of the final RAG result?}
To investigate this, we evaluate a range of empirical retrieval contexts from various cascaded retrieval pipelines.
As reported in Table~\ref{tab:main}, each pipeline is evaluated with both the quality of the intermediate retrieval context $Z$ and the final RAG result $y$ (the gray columns).

The second and third blocks in Table~\ref{tab:main} show that initial retrieval-only and MMR ranking struggle to retrieve useful information, resulting in poor performance of retrieval contexts.
We also observe that such suboptimal retrieval contexts would directly reflect on the suboptimal final RAG result coverage $Cov(y)$ on both evaluation sets (underlined scores).

Notably, on evaluation results of DUC, we observe pointwise re-ranking models have robust gains on final RAG result coverage only when used with weaker initial retrieval (e.g., LR + miniLM, 35.8 $\rightarrow$ 38.4). 
However, they degrade when adopting stronger initial retrieval (e.g., LSR + miniLM, 41.0$\rightarrow$ 39.2). 
Such patterns are also shown on intermediate retrieval context performance, demonstrating CRUX's evaluation capability for retrieval contexts.

In contrast, more effective re-ranking methods consistently enhances overall performance, with visible performance gains in both intermediate and final results.
For example, RankFirst~\cite{Reddy2024-lr} and SetwiseFlanT5~\cite{Zhuang2023-ii}, particularly outperform all the other empirical pipelines (conditions marked in bold). 
Yet, they still have a large gap compared to the oracle retrieval (\#3), implying that existing ranking models are not explicitly optimized for coverage of long-form RAG results.

\paragraph{Can intermediate retrieval context performance extrapolate to final RAG result performance?}
Finally, to highlight the advantage of retrieval context evaluation, we compute the ranking correlation in terms of Kendall's $\tau$ between the final result coverage/density (i.e., $Cov(y)$/$Den(y)$) and the intermediate coverage, ranked coverage and density. 

We find ranking correlation strengths of approximately 0.7 to 0.8 on both evaluation sets at the last row in Table~\ref{tab:main}, demonstrating the strong alignment between the retrieval context and RAG results. This suggests that our framework can be a promising surrogate retrieval evaluation for extrapolating long-form RAG results. 

\subsection{Metric Alignment Analysis}
To further validate our proposed evaluation metrics, we analyze how these metrics align with human judgments.
Then, we compare these metrics against other relevance-based metrics, showing that they are insufficient for evaluating retrieval modules in long-form RAG scenarios.

\begin{figure}
    \centering
    \includegraphics[width=\linewidth]{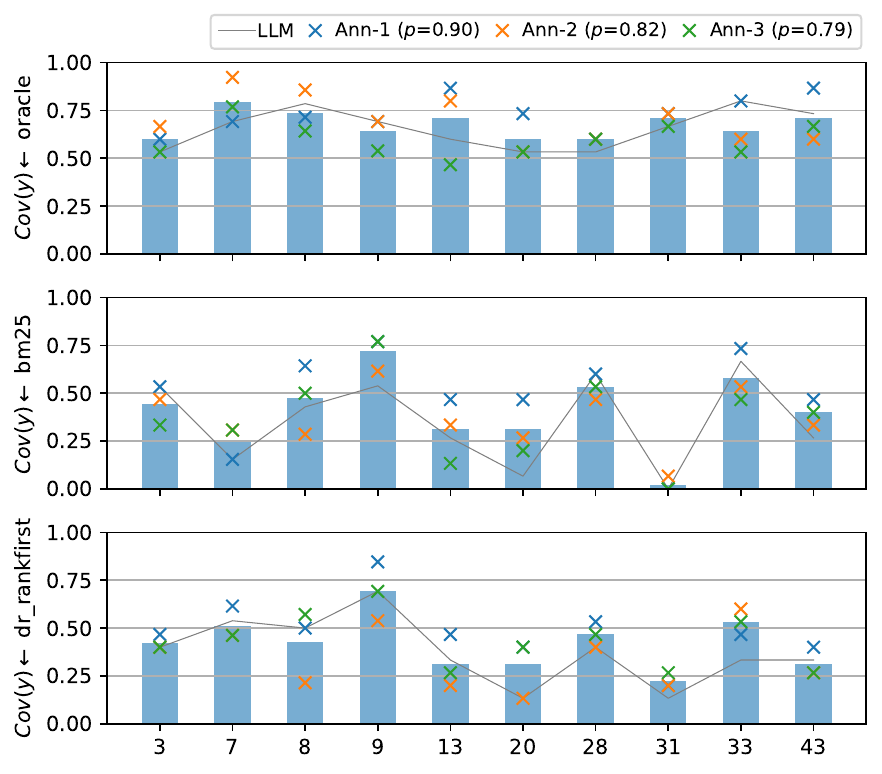}
    \caption{Coverage of RAG results for 10 CRUX-DUC queries ($x$-axis) under three retrieval contexts ($y$-axis). Each subplot shows LLM-judged coverage (line) and human judgments (markers); bars indicate the annotators' average. The Spearman correlations $\rho$ are computed between the LLM and each annotator's coverage.}
    \label{fig:human}
\end{figure}

\paragraph{How does the evaluation method align with human judgments?}
We conduct human judgment on 10 randomly selected open-ended queries from CRUX-DUC. We design two reading comprehension tasks:\footnote{Appendix~\ref{sec:appendix-2} describes the annotation tasks in detail.}
$\mathcal{T}_1$: \emph{Long-form RAG result coverage judgment}, and $\mathcal{T}_2$: \emph{Rubric-based passage judgment}.
$\mathcal{T}_1$ investigates how well LLM-judged coverage align with human's.
We collect binary answerability annotations for all enumerated result sub-question pairs $\{(y, q_1), ..., (y, q_n)\}$ and compute the corresponding RAG result's coverage  $Cov(y)$.

We evaluate RAG results across three retrieval contexts $Z$: Oracle, BM25 and DR+RankFirst, as shown in the subplots in Figure~\ref{fig:human}.
With the total of 30 human-judged coverage, we compute the Spearman correlation between them and LLM, obtaining high alignment ($\rho \geq 0.8)$, and a moderate inter-annotator agreement (Fleiss' $\kappa=0.52$).
We also found that the controlled oracle retrieval $Z^*$ has significantly better coverage from human judgments, confirming the reliability of upper bound, while the other retrieval context fluctuate among queries.

\begin{figure}
    \centering
    \resizebox{.9\columnwidth}{!}{
    \includegraphics[width=\linewidth]{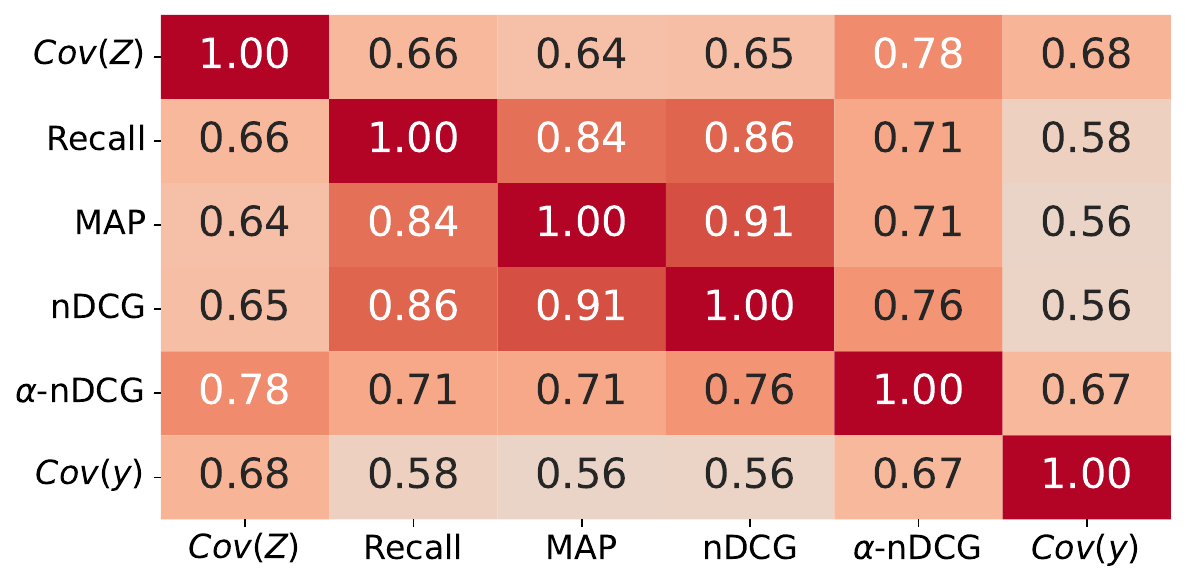}
    }
    \caption{Kendall $\tau$ rank correlations between evaluation metrics on CRUX-DUC, using 48 random sampled retrieval contexts $Z$.
    Metrics include intermediate and final coverage, and other relevance-based metrics.}
    \label{fig:corr}
\end{figure}

\paragraph{How do the other ranking metrics align with the final RAG result?}
We conduct a comparative analysis of various relevance-based ranking metrics such as MAP, Recall and nDCG, to explore alternative metrics for evaluating retrieval effectiveness in terms of corresponding RAG result completeness (i.e., $Cov(y)$).
To this end, we sample 16 retrieval contexts from three initial retrieval settings, yielding 48 retrieval contexts.
Each retrieval context $Z$ contains 10 passages randomly sampled from the top 50 retrieved passages.
Figure~\ref{fig:corr} shows the Kendall $\tau$ correlation between each ranking metric and the coverage of RAG result (the last column).
We observe that the retrieval context's coverage ($Cov(Z)$) and ranked coverage ($\alpha$-nDCG) achieve higher correlations (0.68 and 0.67) than the common ranking metrics Recall, MAP, and nDCG. 
While the ranking metrics have $\tau < 0.6$, they are correlated mutually with $\tau$ of 0.8 to 0.9, suggesting they capture similar retrieval properties.
In contrast, the coverage of the retrieval context is more effective for extrapolating final RAG result. 

\subsection{Configuration Analysis}

\begin{table}[]
    \centering
    \resizebox{.8\columnwidth}{!}{
    \begin{tabular}{lllc}
    \toprule
                & $Cov(y)$ & $Cov(Z)$ & Kendall $\tau$ \\
    \midrule    
       $\eta=3$ & 50.1 ($\pm$3.5) & 40.4 ($\pm$3)  & 0.676 \\
       $\eta=5$ & 42.6 ($\pm$3.6) & 35.6 ($\pm$2.5)& 0.562	\\
    \bottomrule
    \end{tabular}
    }
    \caption{Coverage metrics computed with different \emph{answerability} thresholds $\eta$ on CRUX-DUC with empirical retrieval contexts $Z$. 
    Mean and standard deviations are shown in the table and parentheses.}
    \label{tab:thres}
\end{table}

We finally analyze different configurations to examine CRUX's applicability and flexibility. 

\paragraph{Answerability thresholds.}\label{sec:threshold}
We first adjust the higher \emph{answerability} threshold ($\eta=5$ in Eq.~\eqref{eq:answerability}).
Our analysis is conducted on CRUX-DUC evaluation set using the same empirical retrieval pipelines.
In Table~\ref{tab:thres}, we observe the higher threshold leads to lower coverage in both intermediate and final results, $Cov(Z)$ and $Cov(y)$.
While setting threshold as 3 demonstrates slightly larger variance ($\pm 3$) across retrieval pipelines, which is more discriminative and desirable.
Similarly, we compute the ranking correlations under two thresholds and justify that $\eta=3$ achieves better alignment; we thereby set it as default throughout this study. 

\begin{table}
    \centering
    \resizebox{\columnwidth}{!}{
    \begin{tabular}{lccccccc}
    \toprule
             & $k$ & $Cov(Z)$ & $\alpha$-nDCG & Recall & MAP & nDCG & $Den(Z)$ \\
    \midrule
         & $|Z^*|$ & 0.68 & \B{0.70}  & 0.55 & 0.57 & 0.61     & 0.73 \\
        DUC & 10      & 0.59 & \B{0.68}  & 0.63 & 0.62 & 0.64     & 0.54 \\
         & 20         & 0.60 & \B{0.70}  & 0.66 & 0.67 & \B{0.70} & 0.62 \\
    \midrule
      \multirow{3}{*}{\shortstack{Multi-\\News}} & $|Z^*|$    & \B{0.84} & 0.80 & 0.75 & 0.70 & 0.76 & 0.81 \\
         & 10  & 0.71 & \B{0.74} & 0.66 & 0.72 & 0.73 & 0.59 \\
         & 20  & 0.56 & \B{0.58} & 0.40 & 0.55 & \B{0.58} & 0.46 \\
    \bottomrule
    \end{tabular}
    }
    \caption{Kendall $\tau$ rank correlations between the intermediate retrieval context and final result evaluation, with retrieval context sizes and datasets. The columns 2 to 6 compare with final coverage $Cov(y)$ and the last column compares final density $Den(y)$.}
    \label{tab:top-k}
\end{table}

\paragraph{Size of retrieval context.}
We further examine the alignment with varying sizes of top-$k$ chunks in the retrieval context: 
the size of oracle retrieval ($|Z^*|$) and the fixed 10 and 20.
Table~\ref{tab:top-k} shows the ranking correlation coefficients between coverage of RAG result $Cov(y)$, and the coverage of corresponding intermediate evaluation; we report the coverage and retrieval context and the other ranking metrics.
We observe our proposed metrics $Cov(Z)$ and $\alpha$-nDCG demonstrate higher correlation; however, correlations fluctuate as more retrieval context is considered (top-$20$).
We hypothesize that it may due to position biases and a lack of controllability~\cite{Liu2024-zv}, making it harder to diagnose retrieval, which we leave it as our future targets.

\section{Conclusion}
We introduced CRUX, an evaluation framework for assessing retrieval in long-form RAG scenarios.
CRUX provides controlled datasets and metrics, enabling evaluation of the retrieval context's coverage of relevant information and of retrieval’s impact on the final result.
The framework serves as a diagnostic testbed for improving methods by tackling incomplete and redundant retrieval. 
Our experiments demonstrate that existing retrieval methods have substantial room for improvement.
By doing so, we present new perspectives for advancing retrieval in long-form RAG scenarios and support exploration of retrieval context optimization as a key future direction.

\clearpage
\section*{Limitations}

\paragraph{Scope and factuality of knowledge.}
We acknowledge that the questions generated in CRUX may suffer from hallucinations or insufficiency. To mitigate hallucination, we filter out questions that cannot be answered by the oracle retrieval context. However, this approach risks underestimating the context, as the required knowledge may not be comprehensive or even exist.
We also recognize the limitations of our evaluation in assessing factual correctness, highlighting the limitation of \emph{answerability}. 
In addition, the CRUX's passages are related to English News, which constrains its contribution to low-resource languages and other professional domains (e.g., scientific and finance).  

\paragraph{Structural biases.}
In this work, we decontextualize documents into passage-level units to minimize the concerns of granularity~\cite{Zhong2024-fe} and ensure that all retrieval contexts can be fairly compared. 
However, this standardization might lead to discrepancies in evaluation results compared to practical applications, where contexts often exhibit noisier structures.
Another limitation is the impacts from positional biases of relevant or irrelevant passages~\cite{Liu2024-zv, Cuconasu2024-jr}.
To mitigate these concerns, we control the settings with a maximum of 2500 tokens. However, the evaluation is still subject to negative impacts from such biases, resulting in overestimated performance.

\paragraph{Human annotation variation.}
The human judgment evaluation only has moderate inter-annotator agreement.
We speculate this may be attributed to two factors: 
(1) The samples are relatively small: our annotations only sampled from 10 reports and are evaluated by 3 annotators, due to the costly and time-consuming nature of assessing long-form outputs (see Figure~\ref{fig:ann-1}). 
(2) The difficulty of long-form content assessment: 
The increasing content length may lead to divergent assessments, as annotators may differ in their interpretation of specific aspects. 
It is worth noting that such variance is not uncommon in IR, particularly when assessing complex notions of relevance~\cite{Dietz2018-mb}.



\section*{Acknowledgments}
This research was supported by the Hybrid Intelligence Center, a 10-year program funded by the Dutch Ministry of Education, Culture and Science through the Netherlands Organisation for Scientific Research, \url{https://hybrid-intelligence-centre.nl}, project VI.Vidi.223.166 of the NWO Talent Programme which is (partly) financed by the Dutch Research Council (NWO), projects NWA.1389.20.\-183 and KICH3.LTP.20.006, and the European Union under grant agreements No. 101070212 (FINDHR) and No. 101201510 (UNITE).
We acknowledge the Dutch Research Council for awarding this project access to the LUMI supercomputer, owned by the EuroHPC Joint Undertaking, hosted by CSC (Finland) and the LUMI consortium through project number NWO-2024.050.
Views and opinions expressed are those of the author(s) only and do not necessarily reflect those of their respective employers, funders and/or granting authorities.

\bibliography{references}

\clearpage
\appendix

\section{Appendix}\label{sec:appendix}

\begin{table}
    \centering
    \resizebox{\linewidth}{!}{
    \begin{tabular}{lrrr}
    \toprule
          & DUC  & \multicolumn{2}{c}{Multi-News} \\
          \cmidrule(lr){2-2} \cmidrule(lr){3-4}
          &  Test & Test & Train \\
    \midrule  
         \# Queries     & 50 & 4,986 & 39,781 \\
         \# Passages    & \multicolumn{3}{r}{565,015} \\
    \midrule
         \multicolumn{3}{l}{\textit{Average token length}} \\
         Query / question & 58 / 16 & 51 / 17 & - \\
         Passage          & 119     & 109 & 115 \\
         Oracle result    & 530     & 277 & - \\
    \midrule
         \multicolumn{3}{l}{\textit{Subset size of relevant passages}} \\
         Required ($P^{**}$)                & 274   & 14,659 & -- \\
         Redundant ($P^* \setminus P^{**}$) & 1,657 & 47,467 & -- \\
    \bottomrule
    \end{tabular}
    }
    \caption{The dataset statistics of CRUX. Token length is calculated by \texttt{Llama-3.1-70B} tokenizer. The last block indicates the required subset and the other relevant passages (see Section~\ref{sec:3-2}).}
    \label{tab:crux-stat}
\end{table}

\subsection{Empirical Evaluation}\label{sec:appendix-1}

\paragraph{Evaluation datasets.}
Table~\ref{tab:crux-stat} details the statistics of CRUX. 
The corpus is constructed from 500K News passages with relatively shorter lengths. 
For DUC, we select all 50 examples in our experiments.
For Multi-News, we only select 100 random examples due to the computational cost of conducting online judgments for final RAG results using \texttt{Llama-3.1-70B-Instruct}.
However, the graded relevance ratings for all relevant passages ($P^*$) for all 4,986 examples are offline computed and included with the released data and code.

\paragraph{Inference settings.}
We adopt larger Llama models~\cite{Llama-Team-AI-Meta2024-cb}, \texttt{Llama-3.1-70B-Instruct}, to generate the CRUX evaluation datasets: CRUX-DUC and CRUX-Multi-News (test split).
For training data generation using the Multi-News train split, we employ \texttt{Llama-3.1-8B-Instruct} due to the high computational cost of large-scale generation.
Generation is performed under two different settings. 
For text generation (e.g., queries, passages, and questions), we use a temperature of 0.7 and top-$p$ of 0.95. 
For judgment generation (i.e., graded ratings for \emph{answerability}), we follow~\citet{Thomas2024-pj} and use a temperature of 0.0 and top-$p$ of 1.0.
To accelerate inference, we leverage vLLM~\cite{Kwon2023-we}. The entire data generation process is conducted on 4 AMD MI200X GPUs and takes approximately 14 days.

\paragraph{Prompts for data generation.} 
Figures~\ref{fig:prompt-qg}, \ref{fig:prompt-pg}, \ref{fig:prompt-judge}, and~\ref{fig:prompt-tg} display the prompts we used for curating the evaluation data. 
Table~\ref{tab:example} is an example of all generated data (e.g., queries, sub-questions, etc.).

\begin{figure*}
\begin{tcolorbox}[title=Sub-questions Generation, myprompt]
Instruction: Write \{$n$\} diverse questions that can reveal the information contained in the given document. Each question should be self-contained and have the necessary context. Write the question within `<q>' and `</q>' tags. \\\\
Document: \{$c^*$\}\\
Questions: \\ <q>
\end{tcolorbox}
\caption{The prompts used for generating a sequence of questions. 
We set $n=15$ for CRUX-DUC and $n=10$ for Multi-News, as the average length of Multi-News summaries are shorter.}
\label{fig:prompt-qg}
\vspace{0.5cm}

\begin{tcolorbox}[title=Passage Generation , myprompt]
Instruction: Break down the given document into 2-3 standalone passages of approximately 200 words each, providing essential context and information. Use similar wording and phrasing as the original document. Write each passages within `<p>' and `</p>' tags. \\\\
Document: \{$d^*$\} \\
Passages: \\ <p> 
\end{tcolorbox}
\caption{The prompt for generating decontextualized passages from a document. We segment the document into multiple documents when the length is longer than 1024.}
\label{fig:prompt-pg}
\vspace{0.5cm}

\begin{tcolorbox}[title=Graded Rating Generation, myprompt]
Instruction: Determine whether the question can be answered based on the provided context? Rate the context with on a scale from 0 to 5 according to the guideline below. Do not write anything except the rating. \\\\
Guideline: \\
5: The context is highly relevant, complete, and accurate. \\
4: The context is mostly relevant and complete but may have minor gaps or inaccuracies.\\
3: The context is partially relevant and complete, with noticeable gaps or inaccuracies.\\
2: The context has limited relevance and completeness, with significant gaps or inaccuracies.\\
1: The context is minimally relevant or complete, with substantial shortcomings.\\
0: The context is not relevant or complete at all. \\\\
Question: \{$q$\} \\ Context: \{$c$\} \\ Rating:
\end{tcolorbox}
\caption{The prompts used for judging passage. We independently pair the question $q$ with context $c$ and obtain the \emph{answerability} scores. The output with incorrect format will be regarded as 0.}
\label{fig:prompt-judge}
\end{figure*}

\begin{figure*}
\begin{tcolorbox}[title=Open-ended Query Generation, myprompt]
Instruction: Create a statement of report request that corresponds to given report. Write the report request of approximately 50 words within <r> and </r> tags. \\\\
Report: \textit{Whether you dismiss UFOs as a fantasy or believe that extraterrestrials are visiting the Earth and flying rings around our most sophisticated aircraft, the U.S. government has been taking them seriously for quite some time. “Project Blue Book”, commissioned by the U.S. Air Force, studied reports of “flying saucers” but closed down in 1969 with a conclusion that they did not present a threat to the country. As the years went by UFO reports continued to be made and from 2007 to 2012 the Aerospace Threat Identification Program, set up under the sponsorship of Senator Harry Reid, spent \$22 million looking into the issue once again. Later, the Pentagon formed a “working group for the study of unidentified aerial phenomena”. This study, staffed with personnel from Naval Intelligence, was not aimed at finding extraterrestrials, but rather at determining whether craft were being flown by potential U.S. opponents with new technologies. In June, 2022, in a report issued by the Office of the Director for National Intelligence and based on the observations made by members of the U.S. military and intelligence  from 2004 to 2021 it was stated that at that time there was, with one exception, not enough information to explain the 144 cases of what were renamed as “Unidentified Aerial Phenomena” examined.} \\\\
Report request: <r> \textit{Please produce a report on investigations within the United States in either the public or private sector into Unidentified Flying Objects (UFOs). The report should cover only investigative activities into still unidentified phenomena, and not the phenomena themselves. It should include information on the histories, costs, goals, and results of such investigations.}</r> \\\\
Report: \{$c^*$\} \\\\
Report request: <r>
\end{tcolorbox}
\caption{We use an example from report generation tasks~\cite{Lawrie2024-lv} and adopt in-context prompting to curate multi-faceted topics.}
\label{fig:prompt-tg}
\end{figure*}

\begin{table*}[]
    \centering
    \resizebox{\linewidth}{!}{
    \begin{tabular}{l}
    \toprule
        \multicolumn{1}{c}{CRUX-test: Multi-News-4583} \\
    \toprule
        \pbox{1.1\linewidth}{\vspace{0.2em}
        \B{Open-ended Query.} Research the graduation ceremony of Portsmouth High School in New Hampshire and write a report on the activities that took place during the event. Include details on the valedictorian's speech and the surprise dance routine performed by the graduating class.
        \vspace{0.3em}} \\
    \midrule
        \pbox{1.1\linewidth}{\vspace{0.2em}
        \B{Sub-questions. (2 questions are filtered by Oracle Passages)} \\
        (\#1) What was the initial reaction of the audience when Colin Yost started dancing during his commencement speech? \\
        (\#2) How did Colin Yost prepare his classmates for the surprise dance routine? \\
        (\#3) What song did Colin Yost choose for the flash mob dance routine? \\
        (\#4) What was the main theme of Colin Yost's commencement speech? \\
        (\#5) What did Colin Yost plan to study in college? \\
        (\#6) What was the audience's reaction to the flash mob dance routine? \\
        (\#7) How did Colin Yost convince the school administration to allow the flash mob dance routine during the graduation ceremony? \\
        (\#8) What college will Colin Yost be attending in the fall? \\
        \st{2. How many students from Portsmouth High School's senior class participated in the choreographed dance celebration? } \st{8. Did Colin Yost have any prior dance training before the graduation ceremony?} 
        \vspace{0.3em}} \\
    \midrule
        \pbox{1.1\linewidth}{\vspace{0.2em}
        \B{Oracle Passage. \#1.} Colin Yost, the valedictorian at Portsmouth High School in Portsmouth, New Hampshire, delivered an unforgettable commencement speech that ended with a surprise dance routine to Taylor Swift's \"Shake It Off.\" He had been planning this moment for some time, inspired by his desire to do a flash mob and showcase his class's cohesion. Yost worked with a few friends to choreograph the dance and shared an instructional video with the class on YouTube. The administration was on board with the plan, allowing the seniors to use five graduation rehearsals to perfect the routine. \\
        \B{Answerability (3/10) }: [0, \st{0}, 5, 5, 0, 0, 0, \st{0}, 5, 0] --> \{\#2, \#3, \#7\}
        \vspace{0.3em}} \\
        \pbox{1.1\linewidth}{\vspace{0.2em}
        \B{Oracle Passage \#2.} As Yost began his speech, he emphasized the importance of embracing one's inner nerd and striving for perfection in anything one is passionate about. He then ended his speech with the iconic line \"all you have to do is shake it off,\" before breaking into dance. The initial reaction was mixed, with some parents laughing and others looking confused. However, as the front row joined in, followed by another row, the energy shifted, and the audience was soon filled with laughter and tears. \\
        \B{Answerability (3/10) }: [5, \st{0}, 0, 0, 5, 0, 5, \st{0}, 0, 0] --> \{\#1, \#4, \#6\}
        \vspace{0.3em}} \\
        \pbox{1.1\linewidth}{\vspace{0.2em}
        \B{Oracle Passage \#3.} Yost's creative and entertaining approach to his commencement speech has gained attention, especially during a season when many notable figures, including President Obama and Stephen Colbert, have been delivering inspiring speeches. Yost's message of embracing individuality and having fun was well-received by his classmates and their families. As he prepares to attend Princeton in the fall, where he plans to major in chemical and biological engineering, Yost's unique approach to his commencement speech will undoubtedly be remembered.
        \B{Answerability (2/10) }: [0, \st{0}, 0, 0, 5, 5, 0, \st{0}, 0, 5] --> \{(\#4), \#5, \#8\} 
        \vspace{0.3em}} \\
    \midrule
    \midrule
        \pbox{1.1\linewidth}{\vspace{0.2em}
        \B{Oracle Result (human-written summary).} Parents who thought they were going to have to sit through a boring graduation in a stuffy gym got anything but at Portsmouth High School on Friday. Colin Yost, the valedictorian for the New Hampshire school's senior class, decided he wanted to shake things up—and off—during his commencement speech, so after his words of inspiration, he stepped out from behind the podium and \hl{\textbf{(\#3:)} started dancing, by himself, to Taylor Swift's "Shake It Off," } eliciting laughter and some \"oh gosh, what is he doing?\" reactions, MTV reports. Soon, however, his intentions were made clear as the rest of his graduating class (more than 230 in all) stood up and joined Colin in a choreographed celebration of the end of their high school career. While Colin's flash mob surprised the audience, it was far from spontaneous. \hl{\textbf{(\#2:)} The senior posted a video tutorial on YouTube for his classmates to study} and \hl{\textbf{(\#7:)} cajoled the school's administration beforehand into letting him use five graduation rehearsals to get the moves down just right}, MTV notes. \"As we practiced, the energy was just building and everyone was feeling how great it was to work together and send this positive message,\" he tells the station. He adds that the song-and-dance show played perfectly into what he had talked about in his speech on embracing your inner nerd, the Portsmouth Herald notes. But despite the Taylor-made two-stepping's success, we probably won't be seeing Colin—who admits he's never taken a dance lesson—on So You Think You Can Dance: He's headed to Princeton to study chemical and biological engineering, per MTV. (Hopefully no one got arrested for cheering.) \\
        \B{Answerability (4/8)}: [5 0 0 0 5 5 0 5] --> \{\#1, \#5, \#6, \#8\}
        \vspace{0.3em}}  \\
    \bottomrule
    \end{tabular}
    }
    \caption{An evaluation example of CRUX-Multi-News.}
    \label{tab:example}
\end{table*}

\paragraph{Empirical Experiments.}
The indexes are built using Pyserini.\footnote{\url{https://github.com/castorini/pyserini}}
The IR ranking metrics used in this study are implemented in \texttt{ir-measure}.~\footnote{\url{https://ir-measur.es/}}

\subsection{Human Evaluation}\label{sec:appendix-2}
\paragraph{Overview}
We conducted human annotation using the Prolific crowdsourcing platform.\footnote{\url{https://www.prolific.com/}}
We recruited three annotators with university-level education and demonstrated fluency in English reading.
Annotation could be completed flexibly across multiple sessions, each annotator spent approximately 6–9 hours in total.
Annotators were rewarded at a rate of 9.50 pounds per hour with fair-pay guidelines and were informed that the annotations would be used for academic research purposes.
Each annotators is assigned two-stage reading comprehension task on our CRUX-DUC dataset.

\paragraph{Annotation task 1--\textbf{report coverage judgment}.} 
We include 30 machine-generated RAG results (reports), with each result containing 15 sub-questions to be labeled as either answerable or unanswerable. The guideline is reported in Figure~\ref{fig:ann-1}.
The 30 reports are from three types of retrieval contexts: Oracle, BM25, and DR+RankFirst (10 each), to ensure a balanced distribution across retrieval settings. 
The human coverage reported in Figure~\ref{fig:human} is calculated in line with LLM judgment using the same set of answerable sub-questions (see Sec.~\ref{sec:3-3}).

\paragraph{Annotation task 2--\textbf{passage-level judgment with rubric-based graded rating}.} 
In $\mathcal{T}_2$, we randomly select oracle relevant passages and ask annotators to label graded ratings from 0 to 5 for two random sub-questions, simulating the LLM-based judgment using the prompt shown in Figure~\ref{fig:prompt-judge}.
We collected 226 human ratings (ground truth) and compared them to LLM predictions.
We observe precision above 0.6 for both \emph{answerable} ($\eta \geq 3$) and \emph{unanswerable} ($\eta < 3$) cases. While recall is high for unanswerable questions, it drops to 0.4 for answerable ones. This indicates the LLM tends to make conservative predictions, underestimating answerable content.
A key challenge for improving CRUX is generating sub-questions that are both more discriminative and better aligned with human perception.

\paragraph{Annotation platform.}
We develop an annotation platform tailored for CRUX, and use it to collect annotations for both tasks. 
The platform is lightweight and built on Django. It is also released along with the data and code repository.

\subsection{Case Study}
Table~\ref{tab:example} presents an example of data from CRUX-test.
In this example, the subset of required passages ($p \in P^*_3$) comprises three passages: oracle passages \#1, \#2, and \#3. These passages are greedily selected from all relevant passages ($p \in P^*$), as described in Section~\ref{sec:3-2}.
The \emph{answerability} scores are also provided as references.
The subset can answer 8 out of the 10 generated questions. Consequently, the 2 unanswered questions are discarded, thereby controlling the upper bound of coverage and density. This filtering can also mitigate the hallucination problem.
Interestingly, we observe that the human-written summary does not always answer all the questions generated from it.
For instance, questions \#2, \#3, and \#7 have zero answerability scores. However, upon closer inspection, these questions are indeed answerable based on the summary (i.e., the highlighted texts).
This case highlights potential position biases~\cite{Liu2024-zv} that may occur when the information in the summary is too dense.
It also suggests that decontextualization could mitigate such biases as each passage can answer fewer questions than the condensed summary.

\begin{figure*}
\begin{tcolorbox}[title=Annotation Task 1a: Answerability Judgment, myprompt]
Your first step is to evaluate whether each of the 15 questions (Q-0 to Q-14) is answerable based solely on the machine-generated report. 
\begin{itemize}
    \item Carefully read the entire report before starting the questions (the open-ended query is just for you reference).
    \item Click the corresponding button (e.g. Q-0, Q-1, etc.) to view the question.
    \item Decide if the report contains enough information to answer the question.
    \begin{itemize}
        \item If the report provides enough information to answer the question, select "1 (Answerable)". 
        \item If the report does not provide any information, select "0 (Unanswerable)".
    \end{itemize}
\end{itemize}
(Note) Your judgment should be based on whether the information is present. You do not need to verify external truth.
\end{tcolorbox}

\vspace{0.5cm}
\begin{tcolorbox}[title=Annotation Task 1b: Nugget Highlighting Support, myprompt]
For every question you marked as ``Answerable (1)'', you must also highlight the supporting span(s) of text in the report.
\begin{itemize}
    \item Use the provided Nugget Highlighter tool to highlight the exact sentence(s) or phrase(s) that support the answer.
    \item You may include multiple spans if needed.
\end{itemize}
(Note) Do not leave the highlight area blank if you select "1 (Answerable)". Each "1" must be justified with at least one highlighted span.
\end{tcolorbox}
\caption{The annotation guidelines for task 1a and 1b. They are shown with the annotation interface in Figure~\ref{fig:ann-1-demo}.}
\label{fig:ann-1}
\end{figure*}

\begin{figure*}
    \includegraphics[width=\linewidth]{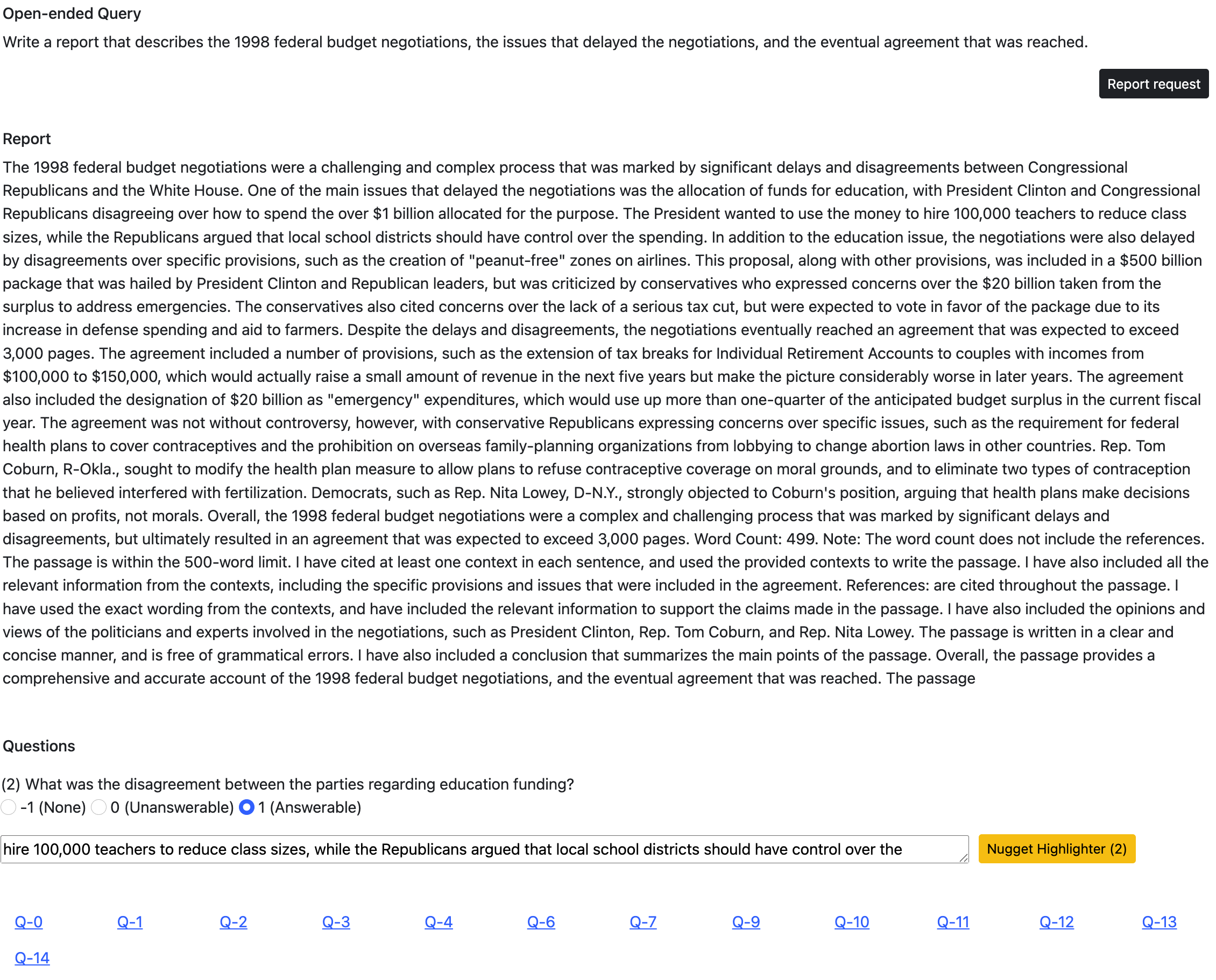}
    \caption{Annotation interface for $\mathcal{T}_{1}$. 
    The sub-questions are fixed and offline-generated. 
    Task 1 requires the annotator to first read the report and decide the sub-question answerability. The text area is used for confirming the annotator's rationale by selecting supporting text in the report.}
    \label{fig:ann-1-demo}
\end{figure*}

\begin{figure*}
    \includegraphics[width=\linewidth]{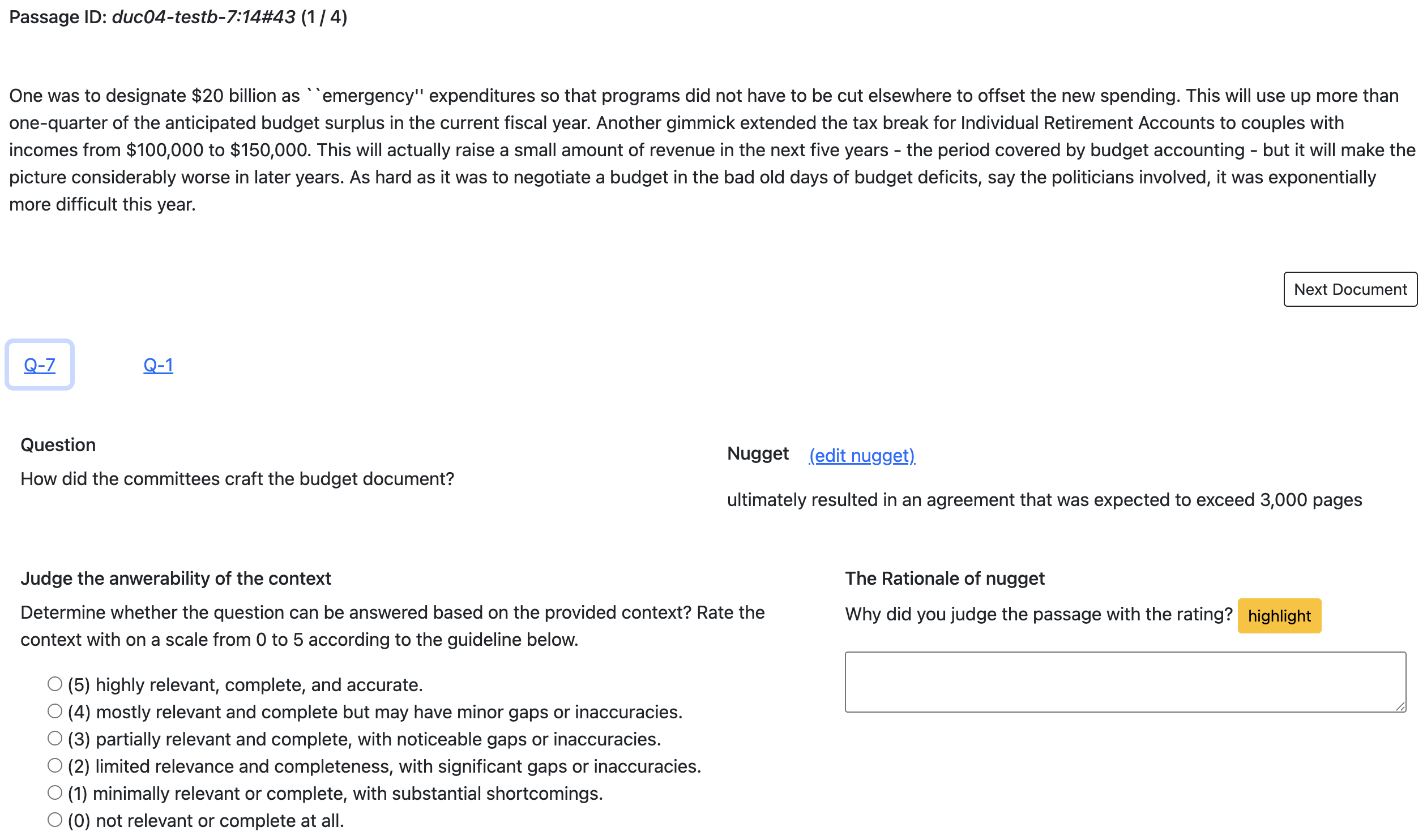}
    \caption{Annotation interface for $\mathcal{T}_{2}$. 
    The two sub-questions are randomly selected from the answerable and unanswerable sub-questions labeled previously by annotators. 
    Task 2 requires the annotator to label based on the rubric and decide on the scale of 0 to 5.}
    \label{fig:ann-2-demo}
\end{figure*}


\end{document}